\documentclass[12pt]{iopart}
\usepackage{pictex}

\usepackage{amsfonts}
\usepackage{amsthm}

\newcommand{\del}{\partial}
\newcommand{\bracket}[2]{\langle #1,#2 \rangle}

\newcommand{\dbyd}[3][]{ \frac{\rmd^{#1} #3}{\rmd #2^{#1}} }
\newcommand{\delbydel}[3][]{ \frac{\del^{#1} #3}{\del #2^{#1}} }
\newcommand{\Lie}{\pounds}
\newcommand{\field}[1]{\mathbb{#1}}
\newcommand{\C}{\field{C}}
\newcommand{\R}{\field{R}}
\newcommand{\N}{\field{N}}

\newcommand{\implies}{\Rightarrow}
\newcommand{\supp}{{\rm supp\ }}

\newcommand{\union}{\cup}
\newcommand{\minus}{\backslash}
\newcommand{\closure}[1]{\overline{#1}}

\newcommand{\definedby}{\stackrel{\rm def}{=} }

\newtheorem{theorem}{Theorem}[section]
\newtheorem{corollary}[theorem]{Corollary}
\newtheorem{lemma}{Lemma}[section]
\newtheorem{definition}{Definition}[section]
\newtheorem{proposition}{Proposition}[section]
\newtheorem{example}{Example}[section]

\newcommand{\eqn}[1]{ \begin{equation} #1 \end{equation} }

\newcommand{\lem}[1]{ \begin{lemma} #1 \end{lemma} }
\newcommand{\defn}[1]{ \begin{definition} #1 \end{definition} }
\newcommand{\propn}[1]{ \begin{proposition} #1 \end{proposition} }

\newcommand{\text}[1]{\mbox{\rm #1 }}
\newcommand{\TFS}{${\cal D}$} 
\newcommand{\tfs}[1]{{\cal D}(#1)}
\newcommand{\DS}{${\cal D}'$} 
\newcommand{\ds}[1]{{\cal D}'(#1)}
\newcommand{\CFS}{${\cal C}$} 
\newcommand{\cfs}[1]{{\cal C}(#1)}
\newcommand{\GFS}{${\cal G}$} 
\newcommand{\gfs}[1]{{\cal G}(#1)}
\newcommand{\SFS}{${\cal C}^\infty$} 
\newcommand{\sfs}[1]{{\cal C}^\infty(#1)}
\newcommand{\mfs}[1]{{\cal C}^\infty_M(#1)}
\newcommand{\nfs}[1]{{\cal C}^\infty_N(#1)}

\newcommand{\Aq}{{\cal A}_q} 
\newcommand{\st}{{\cal M}} 
\newcommand{\assoc}{\approx}
\newcommand{\cphi}{\varphi}
\newcommand{\nn}{\nonumber}

\begin{document}

\title{Signature Changing Space-times and the New Generalised Functions}

\author{Waseem Kamleh}
\address{ Department of Physics and Mathematical Physics\footnote[1]{ADP-00-11/M90}, \\ University of Adelaide, SA 5005, Australia \\ E-mail: wkamleh@physics.adelaide.edu.au }
\date{April 18, 2000}

\submitted

\begin{abstract}
A signature changing spacetime is one where an initially Riemannian manifold with Euclidean signature evolves into the Lorentzian universe we see today. This concept is motivated by problems in causality implied by the isotropy and homogeneity of the universe. As initially time and space are indistinguishable in signature change, these problems are removed. There has been some dispute as to the nature of the junction conditions across the signature change, and in particular, whether or not the metric is continuous there. We determine to what extent the Colombeau algebra of new generalised functions resolves this dispute by analysing both types of signature change within its framework. A covariant formulation of the Colombeau algebra is used, in which the usual properties of the new generalised functions are extended. We find that the Colombeau algebra is insufficient to preclude either continuous or discontinuous signature change, and is also unable to settle the dispute over the nature of the junction conditions.
\end{abstract}

\maketitle

\section{Signature Change in Cosmology}

 The topic of signature change is motivated by problems with causality implied by the observed isotropy and homogenity of the universe. Quantum cosmology\cite{wiltshire} presents us with the possibility that although space-time is currently Lorentzian (i.e. psuedo-Riemannian), the universe may have evolved from an initial state where space-time was Euclidean (i.e. Riemannian) in nature. These so-called signature changing space-times do not possess an initial singularity. Further, as there is now (initially) no distinction between time and space, the problems involving causality are removed.

 Signature changing cosmologies are characterised by a division into a Euclidean region and a Lorentzian region. The two regions are separated by a spatial hypersurface. The difficulties in dealing with a signature changing cosmology arise when one comes to the (delicate) matter of analysing quantities on or across the boundary hypersurface. In general, one obtains junction conditions across the hypersurface by requiring that certain quantities are well-defined. Of course, this will depend on what one requires to be well-defined, and what one means by well-defined. Typically, junction conditions are obtained by requiring continuity of a particular field or its derivatives. In the case of signature change, it would seem that requiring a continuous metric might be a natural condition. However, if we demand that the lapse is continuous, then this requires that it vanishes on the boundary hypersurface, and therefore the metric is degenerate at that point (and thus the inverse metric is singular). On the other hand, if we allow a discontinuous lapse, this requires a distributional metric in order for the derivatives of the metric to be well defined. Whether one of these conditions is better or more natural than the other has been the matter of some dispute. There have been various treatments of the subject, some arguing for continuous (sometimes called {\it strong}) signature change \cite{hayward-sig}, some discontinuous ({\it weak}) signature change \cite{ellis}, and some regarding both as equally valid.

 We note that in both continuous and discontinuous signature change it is possible to analyse the situation within a distributional framework. However, in a distributional framework field equations such as the Klein-Gordon or Einstein equations may contain products and quotients of distributions, which are not well defined. This brings us to the subject of Colombeau algebras, also called the new generalised functions. Within the Colombeau framework, rigorous meaning is giving to non-linear operations on distrib\-utions\cite{colom-mult}, which can be extended to tensor distributions\cite{vickers-wilson}. In signature change, where formal calculations involving distributional products have occurred previously, it has been claimed\cite{hayward-junc} that any difficulties there might be solved by application of the Colombeau algebra. It is our aim to determine the extent to which this is true. We analyse both continuous and discontinuous signature change within the framework of new generalised functions.

 We begin by defining a version of the new generalised functions suited to our needs. Subsequently we analyse continuous signature change, and are forced to develop a means of dividing by generalised functions. To finish, we perform a similar analysis within discontinuous signature change. Continuity conditions are derived in both cases.

\section{Colombeau Algebras \label{ch:colom-alg} }

 Originally, Colombeau\cite{colombeau} developed the space of new generalised functions to deal with products of distributions that occur in quantum field theory. Since then, there have been several variants of the new generalised functions presented. In this section, we construct an algebra of new generalised functions, \GFS, that is based upon one of these presentations. These new generalised functions can be freely summed, multiplied and differentiated. Also, we will show that the smooth functions, continuous functions and distributions can be embedded in \GFS, and their properties are generalised in a consistent fashion. In addition, we construct the field of generalised numbers, where our generalised functions will take point values. The main idea behind \GFS\ is that each element has some ``microscopic'' structure, whose description is lacking in the distributions, which allows us to resolve the ambiguity in multiplication.

\subsection{The algebra of new generalised functions}

 The formulation of \GFS\ given here is based upon a simplified presentation given by Colom\-beau \cite{colom-mult}. It is well known that a distribution can be considered as the limit of a sequence of test functions. In a similar fashion, we will make use of the smooth function space, and define Colombeau objects as an ideal limit of a sequence of smooth functions. As we will perform calculations on curved space time, we require that the formulation of the Colombeau algebra be invariant under general coordinate transformations (diffeomorphisms).\footnote[1]{It has been indicated to us that the covariance of \cite{balasin} (upon which our formulation of \GFS\ is based) may fail. However, we remain in the formalism of \cite{balasin} for reasons of simplicity, and refer the reader to the acknowledgements section at the end of this paper for further clarification. }

Space-time is assumed to be an $n$-dimensional differentiable manifold, $M$, whose tangent bundle is denoted $TM$. The set of sections (vector fields) on the tangent bundle is denoted $\Gamma(TM)$, and the Lie derivative with respect to the vector field $V$ is denoted $\Lie_V$. The elements of the Colombeau algebra, \GFS, are one-parameter families of moderate smooth functions modulo negligible families. For those familiar with other presentations of Colombeau algebras\cite{balasin, colom-mult}, note that we have provided the additional requirement that our families of smooth functions be continuously parameterised.

 \defn{ The space of moderate functions is the set of one-parameter families of smooth functions with continuous parameter $\epsilon \in (0,1]$ defined by  $\mfs{M}=\{(f_\epsilon ) |f_\epsilon \in \sfs{M} \mbox{ such that }\\ \forall \; \ compact\ K \subset M, \forall \;  \{X_1,\ldots,X_p\}, p \geq 0\ with\ X_i \in \Gamma(TM)\ and\ [X_i,X_j]=0, \exists \: N \in {\N}, \exists \:\ \eta > 0, \exists \:\ c > 0,$ such that
 \[\sup_{x \in K} | \Lie_{X_1} \ldots \Lie_{X_p} f_\epsilon(x)| \leq \frac{c}{\epsilon^N}\ for\ 0 < \epsilon < \eta \}.\] }

 \defn{ The space of negligible functions is the set of one-parameter families of smooth functions with continuous parameter $\epsilon \in (0,1]$ defined by  $\nfs{M}=\{(f_\epsilon ) |f_\epsilon \in \mfs{M} \mbox{ such that }\\ \forall \; \ compact\ K \subset M, \forall \;  \{X_1,\ldots,X_p\}, p \geq 0\ with\ X_i \in \Gamma(TM)\ and\ [X_i,X_j]=0, \forall \;  q \in {\N}, \exists \:\ \eta > 0, \exists \:\ c > 0,$ such that
 \[\sup_{x \in K} | \Lie_{X_1} \ldots \Lie_{X_p} f_\epsilon(x)| \leq c \epsilon^q\ for\ 0 < \epsilon < \eta \}.\] }

 \defn{The space of new generalised functions is defined as the ring of equivalence classes of moderate functions modulo the ideal of negligible functions:
 \[ \gfs{M}=\frac{\mfs{M}}{\nfs{M}}. \]}

 Operations in this formulation of \GFS\ are relatively straightforward. Addition, subtraction and multiplication of Colombeau objects are simply defined in terms of the corresponding operations upon their representatives. Multiplication by scalars and partial differentiation are similarly defined in the obvious way.

 We now turn to the embedding of the continuous functions and distributions within \GFS. The embedding is performed by convoluting with a smoothing kernel. The smoothing kernel is an approximate delta-function in order to provide a good generalisation of the classical function product\cite{colom-mult}. Specific use of the tangent bundle is made to preserve diffeomorphic invariance\cite{balasin}.  Given coordinates $\{x\}$ on $M$, we have an induced basis for $TM$ defined by the coordinate derivative fields. Denote this basis by $\{(x,\xi)\}$. The set of test functions is denoted by \TFS\ and the distributions by \DS.

 \defn{(1) Given $q \in {\N}$, we define the set $\Aq = \{ \cphi \in \tfs{{\R}^n}| \int \rmd^n\xi\ \cphi(\xi) = 1,\ and\ \int \rmd^n\xi\ \cphi(\xi)\xi^i=0\ \forall \; \ i \in {\N}^n, with\ 1 \leq |i| \leq q\}.\\$
 (2) Given $\cphi \in \Aq, x \in M$,  we define the function $\cphi_{\epsilon,x} \in \Aq\ by\ \cphi_{\epsilon, x}(\xi)=\frac{1}{\epsilon^n} \cphi(\frac{\xi - x}{\epsilon}).$ }

For details regarding the construction of the set $\Aq$, see \cite{balasin, colom-mult}.
 
 \defn{ Given $f \in \cfs{M}$, we define its associated generalised function $\tilde{f} \in \gfs{M}$ as having a representative $(f_\epsilon)_{0<\epsilon<1}$ where
 \[ f_\epsilon (x) = \int \rmd^n \xi \cphi_{\epsilon,x} (\xi) f(\xi) =
  \int \rmd^n \xi \cphi(\xi) f(x+\epsilon\xi), \cphi \in \Aq. \] }

 \defn{ Given $T \in \ds{M}$, we define its associated generalised function $\tilde{T} \in \gfs{M}$ as having a representative $(T_\epsilon)_{0<\epsilon<1}$ where $T_\epsilon(x)=\bracket{T}{\cphi_{\epsilon,x}}$. If T has a density, g, then
 \[ T_\epsilon (x) \equiv g_\epsilon(x)= \int \rmd^n \xi \cphi_{\epsilon,x} (\xi) g(\xi) =
  \int \rmd^n \xi \cphi(\xi) g(x+\epsilon\xi), \cphi \in \Aq. \] } 

Smooth functions on $M$ are naturally embedded into \GFS\ as constant sequences, $f_\epsilon=f$. But \SFS\ is a subset of continuous function space, \CFS, and thus subject to the above embedding. We would like the two embeddings to coincide. By Taylor-expanding $f$ in the above formula, it is clear that the moment conditions on $\cphi$ will guarantee that the first $q$ non-constant terms vanish. A sufficient condition that the two different embeddings coincide for all smooth functions is requiring that $\cphi \in {\cal A}_\infty$. However, we cannot find such a function\cite{colom-mult}, so rather than choosing a specific smoothing kernel, we allow all functions in $\Aq$, with $q$ arbitrarily large.

Distribution space, \DS, may be extracted from \GFS\ using an equivalence relation on \GFS\ called association.

 \defn{ Given $f,g \in \gfs{M}, if\ \forall \;  \psi \in \tfs{M},$
 \[ \lim_{\epsilon\to 0} \int (f_\epsilon(x)-g_\epsilon(x))\psi(x)\rmd x=0,\] then we say the two generalised functions f and g are associated, denoted $f \assoc g$.}

 \defn{ Given $f \in \gfs{M}$, we set $\bracket{\bar{f}}{\psi} = \lim_{\epsilon \to 0} \int f_\epsilon(x)\psi(x) \rmd x$. If the limit exists $\forall \;  \psi \in \tfs{M}$, then $\bar{f}$ is a distribution defined by this relation. We say that f has an associated distribution, $\bar{f}$. } 

 It is clear that given two associated generalised functions, then if they project onto \DS, they will have the same associated distribution. Furthermore, as an equivalence relation it is clear that the operations of addition, subtraction, derivation and multiplication by scalars are respected by association. That is, acting identically on two associated elements by one of these operations preserves association. Also, multiplication by smooth functions (that do not depend on $\epsilon$) will also preserve association. In this way one can consider association as being equivalent to distributional equality. However, note that multiplication by generalised functions does not preserve association.

 As $\cfs{M} \subset \ds{M}$, the projection of \GFS\ onto \DS\ defines an indirect projection onto \CFS. However, \CFS\ is not a subalgebra of \GFS. So if $f_1,f_2 \in \cfs{M}$, then in \GFS, $\tilde{f_1}\cdot\tilde{f_2} \neq \widetilde{f_1 f_2}$ in general. However, we do have that $\tilde{f_1}\cdot\tilde{f_2} \assoc \widetilde{f_1 f_2}$. This is apparent when one notices that $\lim_{\epsilon \to 0} f_\epsilon(x)=f(x), {\rm for\ } f\in \cfs{M}$. So it is in this way that \GFS\ provides us with a good generalisation of the classical product.

 \subsection{Point values and the generalised numbers.}

 In general, distributions in the classical sense have no natural concept of a point value. For example the Dirac delta function has no classical value at the origin. New generalised functions differ from this in that they have a well defined value associated to a point $x \in M$. However, this will not in general be a classical number, but rather a ``generalised number''.  While the original presentation\cite{colombeau} provided a formal definition for generalised numbers, the presentation which our formulation is based upon\cite{colom-mult} did not. We will define generalised numbers in a similar way to generalised functions.

 \defn{ The space of moderate numbers is the set of continuous one-parameter families of complex numbers defined by  ${\C}_M=\{(z_\epsilon )|z_\epsilon \in {\C} \mbox{ such that } \exists \: N \in {\N}, \exists \:\ \eta > 0, \exists \:\ c > 0,\ such\ that\ | z_\epsilon | \leq \frac{c}{\epsilon^N}\ for\ 0 < \epsilon < \eta \}.$ }

 \defn{ The space of negligible numbers is the set of moderate numbers defined by  ${\C}_N=\{(z_\epsilon )|z_\epsilon \in {\C}_M \mbox{ such that } \forall \;  q \in {\N}, \exists \:\ \eta > 0, \exists \:\ c > 0,\ such\ that\ | z_\epsilon | \leq c \epsilon^q\ for\ 0 < \epsilon < \eta \}.$ }

 \defn{The space of generalised numbers is defined as
 \[ \bar{{\C}}=\frac{{\C}_M}{{\C}_N}. \] }

 \defn{Given a generalised function $f \in \gfs{M},\ and\ x \in M$,we define its point value $f(x) \in \bar{{\C}}$ by $f(x)=(f_\epsilon (x)).$ }
 Using these definitions, it is easy to see that the point value for $f$ is well defined at each point. There are however some further refinements that can be made. In particular ${\C}$ is embedded in $\bar{\C}$ as the set of constant sequences. Similar to the generalised functions, it is possible for a generalised number to have an associated complex number.

 \defn{Given a generalised number $\tilde{z}\in \bar{\C}$, if the limit $z=\lim_{\epsilon\to 0} (z_\epsilon) $ exists in the complex numbers, then we say that $\tilde{z}$ has an associated complex number z.}
 It is clear that this limit is independent of the representative of $z$. The set of all generalised numbers that have an associated complex number form a subalgebra of $\bar{\C}$. Given a generalised function that has an associated complex number at each point, we have a natural projection onto the space of complex functions. 

 \defn{Let $\tilde{f}\in\gfs{\st}$. If $\forall \;  x \in \st$,\ the point value $\tilde{f}(x)$ has an associated complex number $f(x)$ then we say that \~f has an associated complex function f defined by $x\mapsto f(x)$.}
 Not all generalised numbers have an associated complex number, and thus not all generalised functions will have an associated complex function. In particular, we note that the generalised Dirac delta function does not have an associated complex function.

 \propn{If $\delta$ is the one-dimensional Dirac delta function, and $\tilde{\delta}$ is its embedding as a generalised function, then the generalised number $\tilde{\delta}(0)$ has no associated complex number.}
 \begin{proof}
 Given a smoothing kernel $\cphi$, the embedding of the delta function is given by
 \eqn{ \delta_\epsilon (x)=\int \rmd\xi\  \cphi(\xi) \delta(x+\epsilon\xi)=\frac{1}{\epsilon} \cphi(\frac{-x}{\epsilon}). }
 So $\delta_\epsilon(0)=\frac{1}{\epsilon}\cphi(0)$ which is clearly undefined in the limit as $\epsilon \to 0$.
 \end{proof}

 We prove one result that we make use of later on.

 \lem{The generalised number zero is non-invertible. \label{lem:zeronotinv}}
 \begin{proof}
 While this may seem trivial, it is possible for zero to have a nowhere zero representative. However as the representative is negligible, it decays faster than any power of $\epsilon$. So its inverse cannot be moderate. We make these statements formal.

 Suppose that $\exists \: w \in \bar{\C}$ such that $w\cdot z=1$. Then this must also be true for their representatives, so $\exists \: \lambda >0\ \text{ such that } \forall \;  \epsilon,\text{ with }  0 < \epsilon < \lambda,\text{ we have } w_\epsilon z_\epsilon = 1 + n_\epsilon \implies w_\epsilon = (1+n_\epsilon)/z_\epsilon\ $, where $(n_\epsilon) \in \C_N$. Now, $(w_\epsilon)$ is moderate so $\exists \: N \in \N, c >0, \exists \: \eta > 0$ such that
 \eqn{ | w_\epsilon | \leq \frac{c}{\epsilon^N}\text{ for } 0 < \epsilon < \eta. \label{eq:modinv} }
 Now, $(z_\epsilon)$ is negligible, so $\forall \;  q \in \N, \exists \: d>0, \exists \: \zeta > 0$ with
 \begin{eqnarray}
 \nonumber &| z_\epsilon | \leq d {\epsilon^q} \text{ for } 0 < \epsilon < \zeta \\
 \implies &\frac{1}{| z_\epsilon |} \geq \frac{1}{d {\epsilon^q}}\text{ for } 0 < \epsilon < \zeta. \label{eq:neginv}
 \end{eqnarray}
 Choice of $q$ large enough guarantees that (\ref{eq:modinv}) and (\ref{eq:neginv}) cannot both be true, thereby contradicting our assumption that z was invertible.
 \end{proof}

 We note that there are generalised numbers which are infinitesimally small (associated to zero), but are not negligible. This is a feature of the microscopic structure of generalised functions, and is what distinguishes them from the distributions, enabling the algebra to be formed. Having defined new generalised functions, we can now proceed to analyse signature change within this framework.

 \section{Continuous signature change}

 We examine the Klein-Gordon equations in the presence of continuous signature change. We initially work within the framework of distribution space, but find that the junction conditions derived will in general possess some arbitrariness due to the degeneracy of the metric at the boundary hypersurface. An analysis is performed in generalised function space to determine if the arbitrariness is removed in this space. 

 \subsection{The metric and connection coefficients}

 We adopt the metric convention $(+---)$ so that the metric for a signature changing space-time is given by
 \eqn{ \rmd s^2 = \eta(t)\rmd t^2 - h_{ij}\rmd x^i \rmd x^j.}
 We divide spacetime into two regions, $M_+$ and $M_-$, where the former is Lorentzian and the latter is Euclidean. These regions are divided by the spatial hypersurface $t=0$, which we will denote $\Sigma=M\minus(M_+\union M_-) $. Given some quantity $A$, we define its discontinuity $[A]=A|_{\closure{M_+}}-A|_{\closure{M_-}}$.

 Now, as the lapse function $\eta$ is positive on $M_+$ and negative on $M_-$, it must be zero on the spatial hypersurface. By performing a Taylor-expansion about $t=0$, we have $\eta(t)=0+\eta'(0)t + O(t^2)$. Motivated by this we will work in a two dimensional space-time, where in coordinates $(t,x)$ sufficiently close to the boundary hypersurface we have
 \eqn{ \rmd s^2 = t\rmd t^2 - \rmd x^2.}
 So the components of the inverse metric are
 \eqn{ g^{\mu\nu}=\left( \begin{array}{cc} 1/t & 0 \\ 0 & -1 \end{array} \right). } 
 We immediately see that, in function space, the inverse metric is singular on $\Sigma$. We avoid this by moving to distribution space, as it is well-known that $1/t$ as a distribution is well defined.

 As a distribution, division by $t$ is defined by the equation $tS=T$, where $T/t\definedby S$. The general solution to this equation\cite{choquet} is $S_T + a\delta$, where $a$ is an arbitrary constant, and $S_T$ is the distribution defined by
 \eqn{ \bracket{S_T}{\psi}=\bracket{T}{\frac{\psi - \psi(0)\vartheta}{t}} } 
 where $\vartheta$ is fixed and satisfies $\vartheta(0)=1$. We note that the the arbitrariness in choosing $\vartheta$ may be compensated for by an appropriate choice of $a$, so we only have 1-parameter solutions. In the case $T=1$ this defines the distribution $1/t$, denoted $S_1$. Thus $S_1$ satisfies
 \eqn{ \bracket{S_1}{\psi}=\int \rmd t\  \frac{\psi(t) - \psi(0)\vartheta(t)}{t}. }
 It is easily checked that the only non-zero connection coefficient is
 \eqn{
 \Gamma_{00}^0 = \frac{1}{2t}.
 }

 \subsection{The Klein-Gordon field}

 We commence our analysis of the Klein-Gordon field, $\phi$, in distribution space. The Klein-Gordon equation is
 \eqn{(g^{\mu\nu}\nabla_\mu\nabla_\nu-m^2)\phi=0.}
 This implies
 \eqn{
 \frac{1}{t}\delbydel[2]{t}{\phi}-\frac{1}{2t^2}\delbydel{t}{\phi}-\delbydel[2]{x}{\phi}-m^2\phi=0.
 }

 The form of the junction conditions that we derive from this equation will depend on how discontinuous $\phi$ is. We assume that $\phi=\phi_+\theta^+ +\phi_-\theta^-$ where $\theta^+$ is the Heaviside distribution corresponding to $\theta(t)$, $\theta^-$ is the distribution corresponding to $\theta(-t)= 1-\theta(t)$, and $\phi_+, \phi_-$ are smooth with support on $M$. We calculate the required derivatives of $\phi$, denoting $'=\dbyd{t}{}$. 
 \begin{eqnarray}
 \del_t\phi&=& \del_t\phi_+\theta^+ + \del_t\phi_-\theta^- + [\phi]\delta \\
 \del_t^2\phi&=& \del_t^2\phi_+\theta^+ + \del_t^2\phi_-\theta^- + 2[\del_t\phi]\delta + [\phi]\delta'.
 \end{eqnarray}
 So the Klein-Gordon equation becomes
 \begin{eqnarray}
 \nn \fl \frac{1}{t}(\del_t^2\phi_+\theta^+ + \del_t^2\phi_-\theta^- + 2[\del_t\phi]\delta + [\phi]\delta') -\frac{1}{2t^2}(\del_t\phi_+\theta^+ + \del_t\phi_-\theta^- + [\phi]\delta)\\
 \lo -(\del_x^2\phi_+\theta^+ + \del_x^2\phi_-\theta^-) - m^2(\phi_+\theta^+ + \phi_-\theta^-)=0.
 \end{eqnarray}
 The division of a distribution $\alpha$ by $t$ and $t^2$ gives
 \begin{equation}
 \frac{1}{t}\alpha \definedby S_{\alpha}+a_\alpha\delta \qquad \frac{1}{t^2}\alpha \definedby R_{\alpha}+b_\alpha\delta + c_\alpha\delta',
 \end{equation}
 where $R_\alpha$ and $S_\alpha$ are defined by
 \begin{equation}
 \bracket{S_\alpha}{\psi}=\bracket{\alpha}{\frac{\psi-\psi(0)\vartheta}{t}}, \qquad
 \bracket{R_\alpha}{\psi}=\bracket{\alpha}{\frac{\psi-\psi(0)\vartheta-t\psi'(0)\vartheta}{t^2}}
 \end{equation}
 and without loss of generality we choose a specific $\vartheta$ with $\vartheta(0)=1, \vartheta^{(n)}(0)=0$.
 We calculate $R_\alpha, S_\alpha$ in each case, using the definitions and l'Hopital's rule, to get:
 \eqn{
 \bracket{R_\delta}{\psi} = \frac{1}{2}\bracket{\delta''}{\psi} \implies R_\delta = \frac{1}{2}\delta''.
 }
 Similarly, we get
 \begin{equation}
 S_\delta = -\delta', \qquad  S_{\delta'}= -\frac{1}{2}\delta''.
 \end{equation}
 In the case of $\alpha\in\{\theta^+, \theta^-\}$ we have the following equations:
 \begin{equation}
  t^2R_{\theta^{\pm}}-\theta^{\pm}=0, \label{eq:lin-dep-rel} \qquad tS_{\theta^\pm}-\theta^\pm=0.
 \end{equation}
 The Klein-Gordon equation is now, using the above relations,
 \begin{eqnarray}
 \fl \nn \Big(a_{\theta^+}\del_t^2\phi_+ +a_{\theta^-}\del_t^2\phi_- + 2 a_{\delta}[\del_t\phi]+a_{\delta'}[\phi]-\frac{1}{2}(b_{\theta^+}\del_t\phi_+ +b_{\theta^-}\del_t\phi_-+b_{\delta}[\phi]) \Big)\delta \\
 \fl \nn +\Big(-2[\del_t\phi]-\frac{1}{2}(c_{\theta^+}\del_t\phi_+ +c_{\theta^-}\del_t\phi_-+c_{\delta}[\phi]) \Big)\delta'
-\frac{3}{4}[\phi]\delta'' + \Big(t\del_t^2\phi_+-\frac{1}{2}\del_t\phi_+ \\
\fl - t^2(\del_x^2\phi_+ + m^2\phi_+)\Big)R_{\theta^+}
 + \Big(t\del_t^2\phi_--\frac{1}{2}\del_t\phi_- - t^2(\del_x^2\phi_- + m^2\phi_-)\Big)R_{\theta^-} = 0.
 \end{eqnarray}
 As $\delta''$ is linearly independent of the other distributions present, this requires that the discontinuity in $\phi$ vanishes. This is reassuring, as it means the matter field is continuous across the hypersurface.  We note that the restriction of the above equation to $M_\pm$ will simply return the usual Klein Gordon equation. This is because $1/t$ is smooth on $M_\pm$, and thus the distributions $\theta^\pm/t|_{M_\pm}$ are well-defined. At first glance this might seem like $\theta/t$ should be well-defined, but we note that the restriction of a distribution to an open set means that the support of the test functions (which is closed) is in that set, and so test functions on $M^\pm$ vanish a finite distance before the origin, which is why $\theta^\pm/t|_{M_\pm}$ is well defined, but $\theta/t$ is not, as it may act on test functions which have support at the origin.

 Requiring each coefficient to vanish is a possible global solution of this equation, however this is an uninteresting solution as it means that the derivatives of the matter field are identically zero. Further, as $\{ \delta, \delta', \theta^+, \theta-, R_{\theta^\pm}, S_{\theta^\pm} \}$ is a linearly dependent set over the smooth functions, it may not be the only solution. In particular, if $\del_t^2\phi_\pm\propto t$ or $\del_t\phi_\pm\propto t^2$ (or $t$) then other solutions may exist. We investigate this possibility by choosing an (somewhat arbitrary) example.

 We assume that we have $\del_t\phi_\pm=a_\pm t^2 \implies \del_t^2\phi_\pm=2a_\pm t \implies [\del_t\phi]=0$. The fact that the Klein-Gordon equation is satisfied on either side of the boundary implies that the coefficients of $R_{\theta^\pm}$ vanish on either side of the boundary. Equation (\ref{eq:lin-dep-rel}) implies that $\supp R_{\theta^\pm} = \supp \theta^\pm = \closure{M^\pm}$. So the coefficients of $R_{\theta^\pm}$ vanish on their support, which implies that these terms are zero. Together with $[\phi]=[\del_t\phi]=0$, the Klein-Gordon equation simplifies to
 \eqn{
 \fl \Big(a_{\theta^+}\del_t^2\phi_+ +a_{\theta^-}\del_t^2\phi_- -\frac{1}{2}(b_{\theta^+}\del_t\phi_+ +b_{\theta^-}\del_t\phi_-) \Big)\delta - \frac{1}{2}(c_{\theta^+}\del_t\phi_+ +c_{\theta^-}\del_t\phi_-)\delta'=0.
 }
As $t\delta=t^2\delta'=0$, the above equation is trivially satisfied. This means that under our assumptions the global Klein-Gordon equation simply reduces to the Klein-Gordon equation on either side of the boundary. The nature of the assumptions about the derivatives of $\phi$ was sufficiently strong as to preclude needing any further constraints
to satisfy the Klein-Gordon equation. Further, our assumptions require the discontinuity in the first derivative of $\phi$ to vanish, and therefore it is in fact zero at the hypersurface, so the Klein-Gordon field is stationary there. We emphasise that this is a requirement of the assumptions that we made for this particular solution, which were such as to eliminate the arbitrary constants, and were fairly strong. In fact, under the particular set of assumptions above, the Klein-Gordon equation in function space is well defined, so an analysis performed in standard function space might be more constructive.  We consider however, whether moving to generalised function space might remove some of this arbitrariness, without having to revert to such strong assumptions.

 \subsection{Inverting generalised functions}

 While the metric from the previous section is smooth, and is thus included in \GFS, the function $1/t$ is not smooth, and is not included in \GFS. Further, it is singular and hence discontinuous at the origin, so our embedding of continuous functions does not help. In distribution space, we solved the equation $tS=1$ to find the inverse of $t$. The solution to this was $S_1+a\delta$. We investigate to what extent this concept helps us invert the generalised function $t$.

 \defn{Given $f \in \gfs{M}$, we say that f is invertible with (natural) inverse $1/f\definedby g$ if $\exists\ g \in \gfs{M}$ such that $fg=1$. }

 The embedding of the distribution 1/t is $\tilde{S_1} + a\tilde{\delta}$ where
 \eqn{ \delta_\epsilon = \frac{1}{\epsilon}\cphi( \frac{-t}{\epsilon} ), }
 and
 \eqn{ S_{1\epsilon}=\int \rmd\xi\  \frac{\cphi(\xi)-\cphi(-t/\epsilon)\vartheta(t+\epsilon\xi)}{t+\epsilon\xi}. }
 We want an inverse $y\in {\cal G}$ that will satisfy $ty=1$.

 \lem{ $t(\tilde{S_1}+a\tilde{\delta})\neq 1 \in {\cal G}.$ }

 \begin{proof}
Suppose that $t(\tilde{S_1}(t)+a\tilde{\delta}(t))=1$, for some $a$. Then substituting the point value $t=0$ in this equation implies that the generalised number 0 has an inverse, $\tilde{S_1}(0)+a\tilde{\delta}(0)$. But this is a contradiction, as the generalised number zero is not invertible by Lemma \ref{lem:zeronotinv}, thus proving our result.
 \end{proof}

 This proof in fact implies that the equation $ty=1$ has no solution $y$ within the generalised functions, thus negating the possibility that natural inverses might give a means of solution. However, there is another (weaker) possibility. Specifically, as our problem originated in standard function space, we consider that it might be satisfactory if, within the generalised functions, our solutions only hold up to association.

 \defn{Given $f \in \gfs{M}$, we say that f is associatively invertible with (associative) inverse g if $\exists\ g \in \gfs{M}$ such that $fg\assoc 1$. }

 We note that in general neither a natural nor an associative inverse will be unique, in a similar fashion to the division of a distribution by a smooth function. Specifically, given some solution $y_0$ to $ty\assoc 1$, any linear combination of generalised functions $w_i$ satisfying $tw_i\assoc 0 $ can be added to $y_0$ to give other associative inverses. We note that the equivalent is true in the case of division of a distribution by a smooth function (which is a standard method), so we do not consider this a flaw in our definition of the associative inverse.  In order to investigate the extent to which our construction assists us, we seek an associative inverse of $t$. We define the generalised function $\omega$ as having a representative satisfying the following $\forall \;  0 < \epsilon < 1$:
\begin{eqnarray}
\nn \omega_\epsilon(t) = \left\{
 \begin{array}{cc}
  1  & \text{when } \mbox{$|t|\geq \epsilon A$}, \text{ with } \mbox{$A>0$} \text{ fixed}, \\
  0  & \text{when } \mbox{$t=0$},
 \end{array} \right. \\
 \text{and when } t \in (-\epsilon A, \epsilon A) \text{ we have }|w_\epsilon(t)|\leq 1.
\end{eqnarray}
 Further we demand that $\omega^{(n)}_\epsilon(0)$ be independent of $\epsilon$, and we denote this value by $\omega^{(n)}(0)$.

 Then the function $\omega_\epsilon/t$ is easily shown to be moderate. We show that it is smooth by induction. The only place we have to worry about is the origin. By l'Hopital's rule we have
 \eqn{
 \lim_{t\to 0}\dbyd{t}{}{\frac{\omega_\epsilon(t)}{t}}= \lim_{t\to 0}\frac{t\omega'_\epsilon(t)-\omega_\epsilon(t)}{t^2}=  \lim_{t\to 0}\frac{t\omega''_\epsilon(t)+\omega'_\epsilon(t)-\omega'_\epsilon(t)}{2t}=\frac{\omega_\epsilon''(0)}{2}. }
 We assume that $\forall \; n < k$ that there exists a smooth function $f_{\epsilon,k}(t)$ such that the following limit exists and is given by
 \eqn{\lim_{t \to 0} \dbyd[k]{t}{} {\frac{\omega_\epsilon(t)}{t}}=\lim_{t \to 0} \frac{f_{\epsilon,k}(t)}{t}, }
 where $f_{\epsilon, k}$ is smooth. Then 
 \begin{eqnarray}
 \lim_{t \to 0} \dbyd[k+1]{t}{} {\frac{\omega_\epsilon(t)}{t}}&=& \lim_{t \to 0} \dbyd{t}{}\frac{f_{\epsilon,k}(t)}{t} \\
 \nn &=&\lim_{t\to 0}\frac{tf'_{\epsilon,k}(t)-f_{\epsilon,k}(t)}{t^2} \\
 \nn &=&\lim_{t\to 0}\frac{tf''_{\epsilon,k}(t)+f'_{\epsilon,k}(t)-f'{\epsilon,k}(t)}{2t} \\
 &=&\frac{f''_{\epsilon,k}(0)}{2},
 \end{eqnarray}
 so taking $f_{\epsilon,k+1}(t)=t\frac{f_{\epsilon,k}''(t)}{2}$ we complete the proof.

 \begin{figure}[ht]
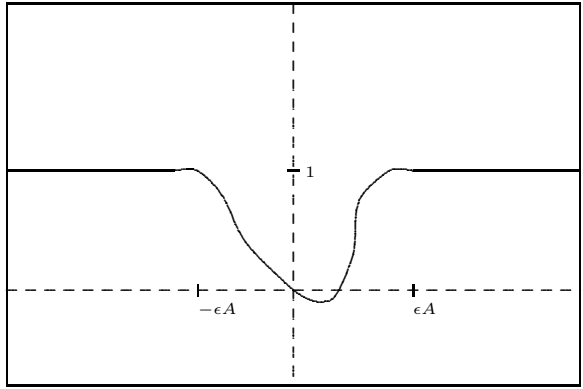

 \centering
 \font\thinlinefont=cmr5
 \begingroup\makeatletter\ifx\SetFigFont\undefined
 \def\x#1#2#3#4#5#6#7\relax{\def\x{#1#2#3#4#5#6}}%
 \expandafter\x\fmtname xxxxxx\relax \def\y{splain}%
 \ifx\x\y   
 \gdef\SetFigFont#1#2#3{%
   \ifnum #1<17\tiny\else \ifnum #1<20\small\else
   \ifnum #1<24\normalsize\else \ifnum #1<29\large\else
   \ifnum #1<34\Large\else \ifnum #1<41\LARGE\else
      \huge\fi\fi\fi\fi\fi\fi
   \csname #3\endcsname}%
 \else
 \gdef\SetFigFont#1#2#3{\begingroup
   \count@#1\relax \ifnum 25<\count@\count@25\fi
   \def\x{\endgroup\@setsize\SetFigFont{#2pt}}%
   \expandafter\x
     \csname \romannumeral\the\count@ pt\expandafter\endcsname
     \csname @\romannumeral\the\count@ pt\endcsname
   \csname #3\endcsname}%
 \fi
 \fi\endgroup
 \mbox{\beginpicture
 \setcoordinatesystem units <0.50000cm,0.50000cm>
 \unitlength=0.50000cm
 \linethickness=1pt
 \setplotsymbol ({\makebox(0,0)[l]{\tencirc\symbol{'160}}})
 \setshadesymbol ({\thinlinefont .})
 \setlinear
 %
 %
 \linethickness= 0.500pt
 \setplotsymbol ({\thinlinefont .})
\putrectangle corners at  2.540 24.130 and 17.780 13.970
 %
 \linethickness= 0.500pt
 \setplotsymbol ({\thinlinefont .})
 \setdashes < 0.1270cm>
\plot  2.540 16.510 17.780 16.510 /
 %
 \linethickness= 0.500pt
 \setplotsymbol ({\thinlinefont .})
 \setsolid
\putrule from  2.540 19.685 to  6.985 19.685
 %
 \linethickness= 0.500pt
 \setplotsymbol ({\thinlinefont .})
\putrule from 13.335 19.685 to 17.780 19.685
 %
 \linethickness= 0.500pt
 \setplotsymbol ({\thinlinefont .})
\putrule from  7.620 16.669 to  7.620 16.351
 %
 \linethickness= 0.500pt
 \setplotsymbol ({\thinlinefont .})
 \setdashes < 0.1270cm>
\plot 10.160 24.130 10.160 13.970 /
 %
 \linethickness= 0.500pt
 \setplotsymbol ({\thinlinefont .})
 \setsolid
\putrule from 13.335 16.669 to 13.335 16.351
 %
 \put{\SetFigFont{6}{7.2}{rm}
$\epsilon A$
} [lB] at 13.335 15.875
 %
 %
 \linethickness= 0.500pt
 \setplotsymbol ({\thinlinefont .})
\putrule from 10.001 19.685 to 10.319 19.685
 %
 \linethickness= 0.500pt
 \setplotsymbol ({\thinlinefont .})
\plot  6.985 19.685  7.008 19.689 /
 \plot  7.008 19.689  7.129 19.706 /
 \plot  7.129 19.706  7.313 19.725 /
 \putrule from  7.313 19.725 to  7.485 19.725
 \plot  7.485 19.725  7.620 19.685 /
 \plot  7.620 19.685  7.747 19.600 /
 \plot  7.747 19.600  7.874 19.488 /
 \plot  7.874 19.488  8.001 19.361 /
 \plot  8.001 19.361  8.128 19.219 /
 \plot  8.128 19.219  8.255 19.050 /
 \plot  8.255 19.050  8.378 18.834 /
 \plot  8.378 18.834  8.490 18.580 /
 \plot  8.490 18.580  8.598 18.307 /
 \plot  8.598 18.307  8.725 18.038 /
 \plot  8.725 18.038  8.890 17.780 /
 \plot  8.890 17.780  9.108 17.520 /
 \plot  9.108 17.520  9.379 17.242 /
 \plot  9.379 17.242  9.671 16.961 /
 \plot  9.671 16.961  9.942 16.705 /
 \plot  9.942 16.705 10.160 16.510 /
 \plot 10.160 16.510 10.327 16.385 /
 \plot 10.327 16.385 10.456 16.311 /
 \plot 10.456 16.311 10.571 16.262 /
 \plot 10.571 16.262 10.681 16.222 /
 \plot 10.681 16.222 10.795 16.192 /
 \plot 10.795 16.192 10.901 16.190 /
 \plot 10.901 16.190 10.996 16.199 /
 \plot 10.996 16.199 11.085 16.216 /
 \plot 11.085 16.216 11.176 16.258 /
 \plot 11.176 16.258 11.271 16.351 /
 \plot 11.271 16.351 11.369 16.512 /
 \plot 11.369 16.512 11.472 16.728 /
 \plot 11.472 16.728 11.576 16.969 /
 \plot 11.576 16.969 11.671 17.215 /
 \plot 11.671 17.215 11.748 17.462 /
 \plot 11.748 17.462 11.794 17.733 /
 \plot 11.794 17.733 11.811 18.034 /
 \plot 11.811 18.034 11.813 18.349 /
 \plot 11.813 18.349 11.836 18.644 /
 \plot 11.836 18.644 11.906 18.891 /
 \plot 11.906 18.891 12.027 19.094 /
 \plot 12.027 19.094 12.190 19.272 /
 \plot 12.190 19.272 12.372 19.435 /
 \plot 12.372 19.435 12.548 19.579 /
 \plot 12.548 19.579 12.700 19.685 /
 \plot 12.700 19.685 12.846 19.734 /
 \plot 12.846 19.734 13.018 19.736 /
 \plot 13.018 19.736 13.197 19.710 /
 \plot 13.197 19.710 13.312 19.689 /
 \plot 13.312 19.689 13.335 19.685 /
 %
 \put{\SetFigFont{6}{7.2}{rm}
1
} [lB] at 10.478 19.526
 %
 %
 \put{\SetFigFont{6}{7.2}{rm}
$-\epsilon A$
} [lB] at  7.620 15.875
 \linethickness=0pt
 \putrectangle corners at  2.515 24.155 and 17.805 13.945
 \endpicture}
 \caption{A possible choice for $\omega_\epsilon(t).$}
 \end{figure}

 \propn{The generalised function $\omega/t$ is an associative inverse of t, and further has an associated distribution which is the principal value of 1/t. }

 \begin{proof}
 Let $\psi \in {\cal D}$ be given. Then  
 \begin{eqnarray}
 \nn \lim_{\epsilon\to 0}\int \rmd t\  (t \frac{\omega_\epsilon}{t}-1)\psi &=&  \lim_{\epsilon\to 0}\int \rmd t\  (\omega_\epsilon-1)\psi \\
 &=&  \lim_{\epsilon\to 0}\int_{-\epsilon A}^{\epsilon A} \rmd t\  (\omega_\epsilon-1)\psi
 \end{eqnarray}
 as $\omega_\epsilon(t)$ is 1 outside of $[-\epsilon A, \epsilon A]$, and as $\omega_\epsilon(t)$ is bounded in this region, this limit clearly converges to zero. So we have $t(\omega/t)\assoc 1$. Now,
 \begin{eqnarray}
 \nn \bracket{P.v.\frac{1}{t}}{\psi}-\lim_{\epsilon\to 0}\int \rmd t\  \frac{\omega_\epsilon}{t}\psi \\
 \fl =\lim_{\delta\to 0}\left( \int_{-\infty}^{-\delta} \rmd t\  \frac{\psi}{t} + \int^{\infty}_{\delta} \rmd t\  \frac{\psi}{t} \right) - \lim_{\epsilon\to 0} \left( \int_{-\infty}^{-\epsilon A} \rmd t\  \frac{\psi}{t} +\int^{\infty}_{\epsilon A} \rmd t\  \frac{\psi}{t} + \int_{-\epsilon A}^{\epsilon A} \rmd t\  \frac{\omega_\epsilon}\psi \right),
 \end{eqnarray}
 which is also clearly zero, thus providing our result.
 \end{proof}

 Before we proceed we note that $t^2$ has an associative inverse given by $\rho/t^2$, where $\rho$ satisfies the same conditions as $\omega$, but also has the property that $\rho'(0)=0$. We also note that as association does not preserve multiplication, the division of each generalised function $f$ by $t$ will be defined by the solution $g$ to the equation $tg\assoc f$.

 \subsection{The Klein-Gordon field revisited}

 We examine the calculations of the previous section, but this time we work in generalised function space. In this section we assume that all references to distributions are taken to mean their embedding within \GFS. The Klein-Gordon equation is formally identical to the previous section

 \begin{eqnarray}
 && (g^{\mu\nu}\nabla_\mu\nabla_\nu-m^2)\phi=0 \\
 \implies && \frac{1}{t}\delbydel[2]{t}{\phi}-\frac{1}{2t^2}\delbydel{t}{\phi}-\delbydel[2]{x}{\phi}-m^2\phi=0.
 \end{eqnarray}

 Once again, we assume that the Klein-Gordon field might be discontinuous in function space. So we have $\phi=\phi_+\theta^+ +\phi_-\theta^-$ where $\theta^+$ is the embedding of the Heaviside function $\theta(t)$, $\theta^-$ is the embedding of the Heaviside function $\theta(-t)= 1-\theta(t)$, and $\phi_+, \phi_-$ are smooth with support on $M$. The derivatives of $\phi$ as a generalised function are as in the previous section: 
 \begin{eqnarray}
 \del_t\phi&=& \del_t\phi_+\theta^+ + \del_t\phi_-\theta^- + [\phi]\delta \\
 \del_t^2\phi&=& \del_t^2\phi_+\theta^+ + \del_t^2\phi_-\theta^- + 2[\del_t\phi]\delta + [\phi]\delta'.
 \end{eqnarray}
 Substituting these into the generalised Klein-Gordon equation gives
 \begin{eqnarray}
 \nn \frac{1}{t}(\del_t^2\phi_+\theta^+ + \del_t^2\phi_-\theta^- + 2[\del_t\phi]\delta + [\phi]\delta')-\\
\frac{1}{2t^2}(\del_t\phi_+\theta^+ + \del_t\phi_-\theta^- + [\phi]\delta)-\del_x^2\phi-m^2\phi=0.
 \end{eqnarray}

 Before we can proceed with the formal divisions we need the following lemma.

 \lem{As generalised functions, $\omega\theta^\pm\assoc \theta^\pm$. }
 \begin{proof}
 For simplicity we assume that $\supp \cphi(-t/\epsilon) = \supp \omega = [-\epsilon A, \epsilon A]$. Then,
 \begin{eqnarray}
 \fl   \lim_{\epsilon \to 0} \int \rmd t\  \theta^+_\epsilon (t) \omega_\epsilon(t)\psi(t) \\
 \nn \fl = \lim_{\epsilon \to 0} \int_{-\epsilon A}^{\epsilon A} \rmd t\  \theta^+_\epsilon (t) \omega_\epsilon(t)\psi(t)
 +\int_{-\infty}^{-\epsilon A} \rmd t\  \theta^+_\epsilon (t) \omega_\epsilon(t)\psi(t)
 +\int_{\epsilon A}^{\infty} \rmd t\  \theta^+_\epsilon (t) \omega_\epsilon(t)\psi(t) \\
 \nn \fl = \lim_{\epsilon \to 0} \int_{-\epsilon A}^{\epsilon A} \rmd t\  \theta^+_\epsilon (t) \omega_\epsilon(t)\psi(t)
 +\int_{\epsilon A}^{\infty} \rmd t\  \theta^+_\epsilon (t) \omega_\epsilon(t)\psi(t),\text{ as $\theta^+$ is 0 on $(-\infty, -\epsilon A),$ } \\
 \nn \fl = \lim_{\epsilon \to 0} \int_{-\epsilon A}^{\epsilon A} \rmd t\  \theta^+_\epsilon (t) \omega_\epsilon(t)\psi(t)
 +\int_{\epsilon A}^{\infty} \rmd t\  \psi(t),\text{ as $\omega, \theta^+$ are 1 on $(\epsilon A, \infty),$} \\
 \nn \fl = \lim_{\epsilon \to 0} \int_{\epsilon A}^{\infty} \rmd t\  \psi(t), \text{ as $\omega, \theta^+$ are bounded on $(-\epsilon A, \epsilon A),$ } \\
\fl = \int_{0}^{\infty} \rmd t\  \psi(t).
 \end{eqnarray}
 Thus we have shown that $\omega\theta^+\assoc \theta^+$. A similar proof applies for $\theta^-$.
 \end{proof}

 It is easily checked from the definition of association that $t\delta \assoc 0$ and $t^2\delta' \assoc 0$. Using this and the above lemma, one can obtain the following

 \begin{eqnarray}
  \frac{\theta^+}{t}\definedby \frac{\omega}{t}\theta^+ +a_{\theta^+}\delta \qquad \frac{\theta^+}{t^2}\definedby \frac{\rho}{t^2}\theta^++b_{\theta^+}\delta + c_{\theta^+}\delta' \\
  \frac{\theta^-}{t}\definedby \frac{\omega}{t}\theta^- +a_{\theta^-}\delta \qquad \frac{\theta^-}{t^2}\definedby \frac{\rho}{t^2}\theta^-+b_{\theta^-}\delta + c_{\theta^-}\delta' \\
  \frac{\delta}{t}\definedby -\delta'+a_{\delta}\delta  \qquad \frac{\delta}{t^2}\definedby \frac{1}{2}\delta''+b_{\delta}\delta + c_{\delta}\delta' \\
  \frac{\delta'}{t}\definedby -\frac{1}{2}\delta''+a_{\delta'}\delta.
 \end{eqnarray}

 Substituting these definitions in, the Klein-Gordon equation becomes
 \begin{eqnarray}
\fl \nn \Big(a_{\theta^+}\del_t^2\phi_+ +a_{\theta^-}\del_t^2\phi_- + 2 a_{\delta}[\del_t\phi]+a_{\delta'}[\phi]-\frac{1}{2}(b_{\theta^+}\del_t\phi_+ +b_{\theta^-}\del_t\phi_-+b_{\delta}[\phi]) \Big)\delta \\
\fl \nn  +\Big(-2[\del_t\phi]-\frac{1}{2}(c_{\theta^+}\del_t\phi_+ +c_{\theta^-}\del_t\phi_-+c_{\delta}[\phi]) \Big)\delta'
 +\del_t^2\phi_+\frac{\omega\theta^+}{t} +\del_t^2\phi_-\frac{\omega\theta^-}{t}\\ 
\fl -\frac{1}{2}(\del_t\phi_+\frac{\rho\theta^+}{2t^2}+\del_t\phi_-\frac{\rho\theta^-}{2t^2}) 
 -\frac{3}{4}[\phi]\delta''-(\del_x^2\phi_+ + m^2\phi_+)\theta^+ -(\del_x^2\phi_- + m^2\phi_-)\theta^- \assoc 0.
 \end{eqnarray}

 Notice that as we have defined the above equation using associative inverses, we can only really consider it valid up to association. Once again we have by linear independence that the Klein-Gordon field must be continuous. What we have done is derived the precise equivalent of the corresponding distributional equation. Thus, it has not helped us remove the arbitrariness in any way, as the linear dependence relations that were present earlier remain. Specifically in this case they are $t\frac{\omega\theta^\pm}{t}\assoc \theta^\pm$ and $t^2\frac{\rho\theta^\pm}{t^2} \assoc \theta^\pm$. So our assertions in the distributional analysis still apply.

 We do however note that using the associative inverses we have defined a division by smooth functions that is consistent with the distributional procedure. The reason for this consistency is that association respects multiplication by smooth functions. While not surprising, this consistency is reassuring, as our definition of the associative inverse reproduces the behaviour seen in the distributions, as would be expected.

We conclude that while it is possible to perform a consistent analysis of continuous signature change within the generalised functions, an analysis performed in function space would probably be more constructive.

Having defined the associative inverse, we are now prepared to analyse discontinuous signature change within the framework of the generalised functions.

 \section{Discontinuous Signature Change}

 A simple model of discontinuous signature change is examined within the framework of the generalised functions. Junction conditions are derived, and it is shown that the Klein-Gordon equation must be satisfied on either side of the boundary hypersurface.

 \subsection{The Klein-Gordon field on a flat background}

 We begin by looking at a simple model of discontinuous signature change. Once again we divide space-time into two disjoint open regions, $M_\pm$, divided by the boundary hypersurface $\Sigma$ defined by $t=0$.  The space-time we will use will be essentially flat, with a discontinuity in the metric at $\Sigma$. The metric is given by
 \eqn{ g = \sigma(t)\rmd t^2 - \rmd x^2 ,}
 where $\sigma$ is the discontinuous function defined by $\sigma(t)=\theta(t)-\theta(-t)$. We can see that this space-time will be flat on both sides of $\Sigma$,and will be Riemannian on $M_-$ and Lorentzian on $M_+$. We note that in order to calculate the connection coefficients, we need to differentiate the metric. This immediately forces us to move to distribution space, as the metric is discontinuous and thus non-differentiable (in function space) on $\Sigma$. We then note we have further difficulties, as the inverse metric is given by
 \eqn{ g^{-1} = \frac{1}{\sigma(t)}\rmd t^2 - \rmd x^2 ,}
 and division by distributions is not defined. We attempt to resolve this by moving to generalised function space.

 The embedding of $\sigma$ in \GFS\ has representative given by
 \eqn{\sigma_\epsilon(t)=\int_{-t/\epsilon}^\infty \rmd\xi\  \cphi(\xi) - \int^{-t/\epsilon}_{-\infty} \rmd\xi\  \cphi(\xi). }
 The first thing to resolve is whether this is invertible within the generalised functions.

 \propn{The generalised function $\tilde{\sigma}$ has no natural inverse.}
 \begin{proof}
 Suppose that $\exists \: g \in {\cal G}$ such that $\tilde{\sigma}   g = 1$. Let $\supp \cphi(t)=[-A,A]$. Then we have
 \eqn{\sigma_\epsilon(t)=\int_{-t/\epsilon}^A \rmd\xi\  \cphi(\xi) - \int^{-t/\epsilon}_{-A} \rmd\xi\  \cphi(\xi). }
 Then we clearly have that $\sigma_\epsilon(t)=1 \, \forall \;  t > \epsilon A$ and $\sigma_\epsilon(t)=-1 \, \forall \;  t < -\epsilon A$. As $\sigma_\epsilon$ is smooth, by the Intermediate Value Theorem, we have that $\forall \;  \epsilon > 0, \exists \: t_0 \in (-\epsilon A, \epsilon A)$ such that $\sigma_\epsilon(t_0)=0$. Now, let 1 be represented by $1+n_\epsilon, n_\epsilon \in {\cal C}_N^\infty$. Then, in particular, $n_\epsilon$ will satisfy $\forall \;  q \in \N, \exists \: c >0, \eta >0$ with
 \eqn{ \sup_{t\in [-\epsilon A, \epsilon A]} |n_\epsilon(t)| < c\epsilon^q \text{ when } 0<\epsilon < \eta.}
 So we can choose some $\lambda$ such that $1+n_\epsilon(t)>0\, \forall \;  t\in[-\epsilon A, \epsilon A]$ whenever $0 < \epsilon < \lambda$. But we have that $\sigma_\epsilon(t) g_\epsilon(t)=1+n_\epsilon (t) \, \forall \;  t \in \R$, and in particular in $[-\epsilon A, \epsilon A]$ we have $0=\sigma_\epsilon(t_0) g_\epsilon(t_0) = 1 + n_\epsilon(t_0) > 0$. This is a contradiction, and thus $\tilde{\sigma}$ has no natural inverse.
 \end{proof}

 As we cannot find a natural inverse we look for an associative inverse of $\tilde{\sigma}$.

 \propn{In generalised function space, $\tilde{\sigma}^2\assoc 1$. }

 \begin{proof}
 We have
 \begin{eqnarray}
 \nn && \lim_{\epsilon \to 0} \int \rmd t\  (\sigma_\epsilon(t)^2-1)\psi(t)\\
 \nn &=& \lim_{\epsilon \to 0} \int_{-\epsilon A}^{\epsilon A} \rmd t\  (\sigma_\epsilon(t)^2-1)\psi(t) \\
 &=& \lim_{\epsilon \to 0} \int_{-\epsilon A}^{\epsilon A} \rmd t\  \Big( \big( \int_{-t/\epsilon}^A \rmd\xi\  \cphi(\xi) - \int^{-t/\epsilon}_{-A} \rmd\xi\  \cphi(\xi) \big)^2-1\Big)\psi(t).
 \end{eqnarray}
 Now, $\cphi$ has compact support so the integrand is well-behaved as $\epsilon \to 0$, and as the integration region vanishes, the limit gives zero, thus the result is immediate.
 \end{proof}

 So we have found a particular associative inverse of $\tilde{\sigma}$, which is itself. In addition, an argument similar to the above shows that $\tilde{\sigma}^3\assoc \tilde{\sigma},\tilde{\sigma}^4\assoc 1$, and so on. We now look for generalised functions $g$ satisfying $\tilde{\sigma}  g \assoc 0$ as these will give a more general associative inverse $\tilde{\sigma} + a_og, a_o\in \C$. Differentiating $\tilde{\sigma}^2\assoc 1$, we obtain $2\tilde{\sigma} ({\tilde{\theta^+}}'-{\tilde{\theta^-}}')\assoc 0 \implies \tilde{\sigma} \tilde{\delta}\assoc 0$. We investigate whether this is the most general solution.

 \propn{Suppose $g\in {\cal G}$ has an associated distribution $\bar{g}$, and satisfies $\tilde{\sigma}  g \assoc 0$. Then $\bar{g}$ has support consisting of at most the origin. }

 \begin{proof}
 We have  $\forall \;  \psi \in {\cal D}$
 \begin{eqnarray}
 \fl \nn && \lim_{\epsilon \to 0} \int \rmd t\  (\sigma_\epsilon(t)g_\epsilon(t))\psi(t)=0 \\
 \fl \implies && \lim_{\epsilon \to 0} \left( \int_{-\infty}^{-\epsilon A} \rmd t\  g_\epsilon(t)\psi(t)  +\int^{\infty}_{-\epsilon A} \rmd t\  g_\epsilon(t)\psi(t) +\int_{-\epsilon A}^{\epsilon A} \rmd t\  \sigma_\epsilon(t) g_\epsilon(t)\psi(t) \right) = 0.
 \end{eqnarray}
 Choice of $\psi$ arbitrary with $\supp \psi \subset (-\infty,0)$ implies that the first term must vanish, similarly if $\supp \psi \subset (0,\infty)$ then the second term vanishes, which implies that $\supp \bar{g} \subset \{0\}$ (using the standard definition of the support of a distribution). 
 \end{proof}

 While it is possible there may be other solutions to $\tilde{\sigma}g\assoc 0$, those solutions will not possess an associated distribution, and will therefore be linearly independent (over the smooth functions) of the functions which do possess an associated distribution. As the coefficient of these will contain an arbitrary constant, we do not consider them in our general solution. Now that we are restricting ourselves to the those generalised functions that have an associated distribution, we know that if a distribution's support consists only of the origin, then it must be a linear combination of the delta distribution or one of its derivatives. Differentiating $\tilde{\sigma}^3\assoc \tilde{\sigma}$ quickly convinces us that $\tilde{\delta}'$ is not a solution, so for our purposes it will be sufficient to assume that $\tilde{\sigma}+a\delta$ is the general solution.

 We proceed to analyse the Klein-Gordon equation. As we are now in generalised function space, we write $\tilde{\sigma}$ simply as $\sigma$, and so on. It is important to note that multiplication by generalised functions does not preserve association, so while we may define an associative inverse of a generalised function, we may not (in general) replace that inverse by something that it is associated to. Further, given a general associative inverse of some function, the associative inverse of the square of that function may not be the square of the associative inverse. As an example, while $\sigma(\sigma + a\delta) \assoc 1$ it is not true that $\sigma^2(\sigma+a\delta)^2\assoc 1$. To see this, consider the following: $\sigma^4 \assoc 1 \implies \sigma^3\delta \assoc 0 \implies 6
 \sigma^2\delta^2+\sigma^3\delta' \assoc 0$, so we do not have $\sigma^2\delta^2 \assoc 0$. So we perform the divisions by generalised functions formally, and only substitute in the associative inverses at the last possible moment.

 The only non-zero connection coefficient is again
 \eqn{
 \Gamma_{00}^0 = \frac{1}{2\sigma}\sigma' = \frac{\delta}{\sigma}.
 }

 The Klein-Gordon equation is
 \begin{eqnarray}
 \nn g^{\mu\nu}(\del_\mu\del_\nu\phi-\del_\lambda\phi\Gamma^\lambda_{\mu\nu})-m^2\phi=0 \\
  \implies  \frac{1}{\sigma}(\delbydel[2]{t}{\phi}-\frac{\delta}{\sigma}\delbydel{t}{\phi})-\delbydel[2]{x}{\phi}-m^2\phi=0.
 \end{eqnarray}
 As in the previous section we assume that the Klein-Gordon field may in general be non-smooth, so we have 
 \begin{eqnarray}
 \phi&=& \phi_+\theta^+ + \phi_-\theta^- \\
 \del_t\phi&=& \del_t\phi_+\theta^+ + \del_t\phi_-\theta^- + [\phi]\delta \\
 \del_t^2\phi&=& \del_t^2\phi_+\theta^+ + \del_t^2\phi_-\theta^- + 2[\del_t\phi]\delta + [\phi]\delta',
 \end{eqnarray}
 where once again $\phi_\pm$ are smooth on $M_\pm$ respectively.
 So the Klein Gordon equation is now
 \begin{eqnarray}
 \nn \del_t^2\phi_+\frac{\theta^+}{\sigma} + \del_t^2\phi_-\frac{\theta^-}{\sigma} + 2[\del_t\phi]\frac{\delta}{\sigma} + [\phi]\frac{\delta'}{\sigma} 
 -\del_t\phi_+{\frac{\theta^+\delta}{\sigma^2}} - \del_t\phi_-\frac{\theta^-\delta}{\sigma^2} - [\phi]\frac{\delta^2}{\sigma^2}\\
 -(\del_x^2\phi_+ +m^2\phi_+)\theta^+ -(\del_x^2\phi_- +m^2\phi_-)\theta^- =0.
 \end{eqnarray}

 We note that as $\sigma=\theta^+-\theta^-$, we have that $\sigma\theta^\pm\assoc \pm\theta^\pm$. In addition, it can be shown\footnote[1]{Simply expand $\sigma^2\assoc 1\assoc \theta^+ +\theta^-$.} that $\theta^+\theta^-\assoc 0$, so we also have $\sigma^2\theta^\pm\assoc \theta^\pm$. Using these relations, and $\sigma^n\assoc 1 \text{ or } \sigma$, it is possible to derive the following:
 \begin{eqnarray}
 \theta^\pm\sigma^{-1}=\pm\theta^\pm + a_\pm\delta \qquad \theta^\pm\delta\sigma^{-2}=\pm\frac{3}{2}\delta \\
 \delta\sigma^{-1}=\frac{3}{2}\sigma\delta +a_{\theta^+}\delta \qquad \delta^2\sigma^{-2}= -\frac{3}{2}\sigma\delta' \\
 \delta'\sigma^{-1}=3\sigma\delta'+a_{\theta^-}\delta.
 \end{eqnarray}
 Using these\footnote[2]{We were unable to derive an equation of the form $\sigma^2g\assoc \delta^2$. However the equation $\delta^2\assoc -\frac{1}{2}\sigma\delta'$ suggests that we might be able to define $\delta^2\sigma^{-2}=-\frac{1}{2}\delta'\sigma^{-1}$. We note that even were we unable to do this, linear independence over the smooth functions would still require $[\phi]$ to vanish, and thus the point is somewhat academic, so to speak. }, our Klein-Gordon equation has become
\begin{eqnarray}
 \fl \nn (\del_t^2\phi_+a_+ + \del_t^2\phi_-a_- + 2[\del_t\phi]a_{\theta^+} + [\phi]a_{\theta^-}-\frac{3}{2}\del_t\phi_++\frac{3}{2}\del_t\phi_-)\delta +[\del_t\phi]3\sigma\delta \\
 \fl +[\phi](\frac{9}{2}\sigma\delta')+(\del_t^2\phi_+ -\del_x^2\phi_+ -m^2\phi_+)\theta^+ +(-\del_t^2\phi_- -\del_x^2\phi_- -m^2\phi_-)\theta^- =0.
\end{eqnarray}
 Now, the generalised functions in the above equation are linearly independent over the smooth functions, so their coefficients must vanish. This implies that the Klein-Gordon field is continuous, as is its first time derivative, and that the Klein-Gordon equation is satisfied either side of the boundary. We must however caution against taking the above equation at face value. While having an associative inverse is a necessary condition for a generalised function to have a natural inverse, we have seen that it is not a sufficient condition. We posit that there must be some doubt as to the validity of defining a multiplicative inverse using an equivalence relation that does not respect multiplication. As an example of this doubt, if one considers the above equation as having a corresponding associative equation, then as $\sigma\delta\assoc 0$ we no longer have the requirement that $\del_t\phi$ is continuous.

 While the generalised functions provide a rigorous framework for the formal operation of multiplying distributions, we have seen that this is not true for the operation of division. As is the case with all formal calculations in physics, it would be desirable if there were a rigorous mathematical theory describing the calculations. A field that contained the generalised functions as a subalgebra would be such a theory. However, as the generalised functions do not form an integral domain, such a field may not exist.

 Noting that we are really only interested in what happens at the boundary hypersurface, one might consider evaluating the Klein-Gordon equation at $t=0$ using the generalised numbers. However, the point value of $\delta$ at $t=0$ does not have an associated complex number, and if the point value of $\theta^\pm$ does have an associated complex number, this number will be embedding dependent. These facts make a point value analysis unuseful. 

 We have seen that our calculations performed within the Colombeau algebra required formal division of generalised functions that do not possess a natural inverse. The validity of the associative inverse is questionable, and thus we have stepped over the bounds of rigor. Therefore we must conclude that the analysis of discontinuous signature change within the new generalised functions is on no better foundation than those within other frameworks.

\section{Conclusion}

The topic of signature change is of physical interest as it removes certain problems in standard cosmology. Within this topic, we have seen why being able to multiply distributions is important, and have presented a formulation of the Colombeau algebra. Within this presentation, rigorous meaning is given to products of distributions, and a good generalisation of the classical function product is given.

An analysis of the Klein-Gordon field under continuous signature change was conducted within the framework of the distributions. The degeneracy of the metric on the boundary hypersurface gave arbitrariness to the junction conditions derived, as we were required to divide by smooth functions that were zero at the hypersurface. An attempt was made to remove this arbitrariness by moving to generalised function space. We gave a natural and a weak method of dividing by generalised functions, and found that not every generalised function possesses a natural inverse. Using associative inverses we found that arbitrariness was still present in the junction conditions derived in generalised function space. This makes sense, as our divisions were performed using association, and modulo association, generalised function space looks like the distributions. We did however derive that the Klein-Gordon field must be continuous in both analyses.

We also looked at junction conditions in discontinuous signature change. We performed an analysis of the Klein-Gordon field over a flat signature changing space time. This time divisions by generalised functions (as opposed to smooth functions) were present, and we found that these functions were not naturally invertible. The method of associative inverses was used to derive junction conditions, and it was found that the Klein-Gordon field had to be continuous across the junction. The Klein-Gordon equation on either side of the boundary hypersurface was required to be satisfied as a direct result of the equations derived. Some arbitrariness was present in these equations, due to the weak nature of the associative equivalence.

We found that we had to extend the Colombeau algebra to include the operation of division by generalised functions. This was done because the Colombeau algebra possesses a product, but not a quotient. As the generalised functions can be multiplied, we had a natural means of defining this division. However, as is the case with smooth functions, we found that we could not always divide by a generalised function. As the generalised functions that we were interested in were not naturally invertible, we developed a weak division using the relation of association. While this allowed us to formally divide by generalised functions, it introduced arbitrariness into our equations. In the case of division by a smooth function, we found that the associative division corresponded exactly to the operation of dividing a distribution by a smooth function. This gives credence to the associative division, at least when one is dividing by smooth functions. We cannot be entirely certain that this is true in the case where we had to divide by arbitrary generalised functions.

Our aim was to determine the extent to which the Colombeau algebra enabled us to rigorously analyse signature changing space times. This was motivated by speculation (e.g. Hayward\cite{hayward-junc}) that the Colombeau algebra would solve problems involving non-linear operations, and also by claims (e.g. Mansouri and Nozari\cite{mansouri-sigchange}) that it did. In the cases of both continuous and discontinuous signature change, we found that we were forced to extend the standard formalism of the new generalised functions. In the case of continuous signature change, the generalised functions did not provide any more insight than the distributions, and in the case of discontinuous signature change, we were forced to make use of formal calculations in order to proceed within the framework of the generalised functions. To state the focus of this paper, we asked the Colombeau algebra to provide a framework within which all quantities of interest and the operations performed upon them were rigorously defined. Specifically, we found that our formulation of the Colombeau algebra and the extensions we made to it were inadequate for this task. Clarifying, we again assert that as association does not respect multiplication, we cannot in general replace an associative inverse by something that it is associated to. This fact questions the validity of formal division using association.

The least that we can conclude from this work is that the specific formulation of new generalised functions given here is unable to provide a rigorous framework for the calculations involved. However, given the nature of the difficulties that we came across, we feel that something more can be said. In any signature changing space-time, the main problems arise from the boundary hypersurface, and in particular, the well-definedness of the inverse metric on the hypersurface. Unless a formulation of the Colombeau algebra is developed where a rigorous means of dividing by the desired generalised functions is given, or a means of circumventing this difficulty is presented, then a lack of rigor will plague all such analyses within the generalised functions. Taking all of this into account, we feel that in answer to the question ``Do current formulations of the new generalised functions provide a rigorous framework for analysing signature changing space-times?'' we must conclude in the negative.
\vspace{5 mm}

\ack

I would like to thank David Hartley for his valuable assistance and many discussions contributing to this publication. Thanks also go to the referees of the initial version for their suggestions, and to M. Kunzinger and R. Steinbauer for their correspondence regarding Colombeau algebras and the generalised numbers. In particular, they pointed out that the construction of \GFS\ based upon \cite{balasin} may not satisfy covariance. It is acknowledged that their point is valid, however, we note that other formulations exist which are covariant, and the ideas contained within this paper should be easily translatable to these other formalisms. Further, the covariance of \GFS\ is not central to the results that have been given here, as we remained within a single set of coordinates, and we only considered scalar generalised functions as opposed to generalised tensors and so forth. We therefore chose to remain in the formalism of \cite{balasin} for simplicity.

\bibliographystyle{abbrv}
\bibliography{reference}

\end{document}